\documentclass[conference]{IEEEtran}

\IEEEoverridecommandlockouts  
\usepackage{helvet}         
\usepackage{courier}        
\usepackage{type1cm}        
\usepackage{makeidx}         
\usepackage{graphicx}        
         
\usepackage{multicol}        
\usepackage[bottom]{footmisc}
\usepackage{ifthen}
\usepackage{subfigure}
\usepackage[usenames,dvipsnames]{color}
\usepackage{cite}

\makeindex             

\usepackage{graphics} 
\usepackage{graphicx} 
\usepackage{epsfig} 
\usepackage{times} 
\usepackage{mathtools}  
\usepackage{amssymb}  
\usepackage{amsthm}

\DeclareGraphicsExtensions{.pdf,.png,.jpg,.eps}

\usepackage{subfig}
\usepackage{verbatim}
\usepackage{comment}
\usepackage{algorithm}
\usepackage{algorithmic}
\usepackage{multirow}
\usepackage{setspace}

\DeclareGraphicsExtensions{.pdf,.png,.jpg,.eps}

\def\ba{\begin{array}}
\def\ea{\end{array}}
\newcommand{\beq}{\begin{equation}}
\newcommand{\eeq}{\end{equation}}
\newcommand{\bq}{\begin{eqnarray}}
\newcommand{\eq}{\end{eqnarray}}
\newcommand{\bqn}{\begin{eqnarray*}}
\newcommand{\eqn}{\end{eqnarray*}}
\newcommand{\bee}{\begin{enumerate}}
\newcommand{\eee}{\end{enumerate}}
\newcommand{\bi}{\begin{itemize}}
\newcommand{\ei}{\end{itemize}}

\newcommand{\rr}{{\mathbb{R}}}

\newboolean{showcomments}
\setboolean{showcomments}{true}
\newcommand{\wang}[1]{\ifthenelse{\boolean{showcomments}}
{ \textcolor{red}{(ZW:  #1)}}{}}
\newcommand{\fliu}[1]{\ifthenelse{\boolean{showcomments}}
{ \textcolor{red}{(FL:  #1)}}{}}
\newcommand{\peng}[1]{\ifthenelse{\boolean{showcomments}}
	{ \textcolor{blue}{(PY:  #1)}}{}}

\theoremstyle{definition}
\newtheorem{theorem}{Theorem}
\newtheorem{lemma}[theorem]{Lemma}

\newtheorem{proposition}[theorem]{Proposition}

\theoremstyle{definition}
\newtheorem{definition}{Definition}
\newtheorem{remark}{\textit{Remark}}

\newtheorem{assumption}{Assumption}

\begin{document}
	
\setstretch{1}
\title{\LARGE \bf Toward Distributed Stability Analytics for Power Systems with Heterogeneous Bus Dynamics}

\author{Peng Yang, Feng Liu, Zhaojian Wang, Chen Shen, Jun Yi, Weifang Lin
	\thanks{This work was supported by State Grid Corporation of China Headquarters Science and Technology Project Funding.}
		\thanks{P. Yang, Z. Wang, F. Liu, and C. Shen are with the Department of Electrical Engineering, Tsinghua University, Beijing, 100084, China {\tt\small lfeng@tsinghua.edu.cn}} 
		\thanks{ J. Yi and W. Lin are with China Electric Power Research Institute, Beijing, 100192, China {\tt\small yjung@epri.sgcc.com.cn}}
}


\maketitle

\begin{abstract}
	The stability issue emerges as a growing number of diverse power apparatus connecting to the power system. The stability analysis for such power systems is required to adapt to heterogeneity and scalability. This paper derives a local passivity index condition that guarantees the system-wide stability for lossless power systems with interconnected, nonlinear, heterogeneous bus dynamics. Our condition requires each bus dynamics to be output feedback passive with a large enough index w.r.t. a special supply rate. This condition fits for numerous existing models since it only constrains the input-output property rather than the detailed dynamics. Furthermore, for three typical examples of bus dynamics in power systems, we show that this condition can be reached via proper control designs. Simulations on a 3-bus heterogeneous power system verify our results in both lossless and lossy cases. The conservativeness of our condition is also demonstrated, as well as the impact on transient stability. It shows that our condition is quite tight and a larger index benefits transient stability. 
\end{abstract}


\IEEEpeerreviewmaketitle

\section{INTRODUCTION}
\label{sec:1}

Traditional power grids mainly consist of centralized synchronous generators (SGs). However, with the concern of the environment and the advancement of renewable energy technologies, the dynamical components in power systems are becoming heterogeneous. Moreover, with the development of micro-grid, numerous distributed energy resources are connected to the network in a decentralized manner\cite{wang2018distributed_coping}. The power system is transferring from a centralized homogeneous generation to a decentralized heterogeneous mode. One major challenge in such power systems is the system-wide stability analysis. In this paper, we aim to derive a localized condition to ensure the system-wide stability with special concerns about heterogeneity and scalability.

Traditionally, the stability is assessed by time-domain simulation\cite{Stott_Powersystemdynamic_1979}, eigenvalue analysis\cite{Sastry_Hierarchicalstabilityalert_1980}, and direct methods\cite{Chiang_Directstabilityanalysis_1995}, all of which are from the centralized perspective. These methods may fail in decentralized heterogeneous power systems due to the computational burden, communication failures, and privacy concerns\cite{Wang2019Distributed_I,Wang2019Distributed_II}. 
Thus, researchers turn to develop distributed stability analytics for power systems. 

One approach is based on decomposition of system Jacobian matrix, such as re-constructing the system-wide Jacobian matrix at each agent\cite{Song_DistributedFrameworkStability_2017}, abstracting the interconnection part\cite{Zhang_OnlineDynamicSecurity_2015}, and regarding the effect of interconnection as disturbance\cite{Ilic_standardsmodelbasedcontrol_2012}. However, this approach is limited to linear dynamics and small-signal stability. Other approaches include using 
linear matrix inequalities\cite{Zhang_TransientStabilityAssessment_2016}, sum-of-square technique and vector Lyapunov functions \cite{Kundu_sumofsquaresapproachstability_2015}. These computation-based methods overlook the structure property of the network and, more importantly, may still suffer from the computational burden.

A more favorable approach is based on the concept of \textit{passivity} and \textit{dissipativity}. This concept has been one of the cornerstones of nonlinear control theory since the 1970s and is widely used in the study of interconnected dynamical systems\cite{Bao_ProcessControlPassive_2007,vanderSchaft_L2GainPassivityTechniques_2017}. In the literature of power systems, it is usually combined with the port-Hamiltonian system framework\cite{VanDerSchaft_PortHamiltoniansystemsnetwork_2004} to study the problem of stability\cite{Fiaz_portHamiltonianapproachpower_2013a,Caliskan_CompositionalTransientStability_2014} and controller design\cite{Wang_preserving,Stegink_unifyingenergybasedapproach_2016}. However, in these works, the network is either assumed to be dissipative\cite{Caliskan_CompositionalTransientStability_2014,Stegink_unifyingenergybasedapproach_2016} or is analyzed centrally\cite{Fiaz_portHamiltonianapproachpower_2013a}. Another useful concept is the \textit{passivity index}, which can be extended to more general cases for both passive and non-passive systems\cite{Sepulchre_ConstructiveNonlinearControl_1997,Li_ConsensusHeterogeneousMultiAgent_2019}. Passivity index quantifies the excess or shortage of passivity of a dynamical system. And the shortage can be compensated by the excess of another interconnected system to enforce the closed-loop stability\cite{Sepulchre_ConstructiveNonlinearControl_1997}.

In this paper, we further adopt the passivity index to the case where the supply rate has differential at one port. Leveraging this specialized concept, a distributed condition for power devices in the network is proposed to ensure the system-wide stability. The salient features of our condition are  threefold:
\begin{itemize}
	\item The passivity index of individual bus dynamics can be examined locally, which empowers scalable stability analytics of power systems. 
	\item It is built on the input-output property rather than the detailed dynamical model, which is adaptable to heterogeneous nonlinear bus dynamics.
	\item It motivates an easy way to adjust the parameters of existing controllers locally to meet the stability condition without redesign.
\end{itemize}
For three specific dynamical models, viz. synchronous generator, conventional and quadratic droop-controlled inverters, we show the desired excess of passivity can be obtained via proper control settings. The result is illustrated and verified by a 3-bus power system simulation.


\section{Preliminaries and Formulations}
Consider a network-reduced power system composed of $n$ buses and transmission lines connecting them. It can be abstracted as an undirected graph $\cal G=(\cal V,\cal E)$, where $\cal V$ is the set of buses and $\cal E$ is the set of lines. Each bus in $\cal G$ is associated with a phasor voltage $V_i\angle\theta_i$ and a complex power injection $P_i+\sqrt{-1}Q_i$. $V_i\in\rr_{>0}$ is the magnitude, $\theta_i\in\rr$ is the phase angle, $P_i\in\rr$ and $Q_i\in\rr$ is the real and reactive power injection, respectively. 

\subsection{Modeling Bus Dynamics}

We assume each bus in $\cal G$ is attached to a dynamical power device, including synchronous machine and inverter-interfaced power source or load. We generally call them bus dynamics as they determine a dynamical relation between $V_i,\theta_i$ and $P_i,Q_i$. 

We consider a generic input-output model for bus dynamics as follows. 

\begin{equation}\label{eq:dynamicV}
\left\lbrace \begin{aligned}
\dot{x}_i&=f_{i}(x_i,u_i)\\
y_i&=h_i(x_i)=(\theta_i,V_i)^T
\end{aligned}\right. \;\;\;i\in\mathcal{V}
\end{equation}
where $x_i=\text{col}(\xi_i,\theta_i,V_i)\in\mathcal{X}_i\times\rr\times\rr_{>0}$ is the state variable of bus dynamics $i$, and $\xi_i\in\mathcal{X}_i$ is the auxiliary state variable which includes the heterogeneous dynamics of each component. The input is the power injection at the bus $u_i=(P_i,Q_i)\in\rr^2$. $f_i:\;\mathcal{X}_i\times\rr\times\rr_{>0}\times\rr^2\to\mathcal{X}_\mu\times\rr\times\rr_{>0}$ is a continuously differentiable function. Note that the choice of output signal $h_i$ is actually free, and we specifically select it as $\theta_i$ and $V_i$ here for the simplicity of analysis. 

The generic model \eqref{eq:dynamicV} covers many existing dynamical models in power systems, such as classical synchronous machine\cite{Stegink_unifyingenergybasedapproach_2016}, inverter-interfaced power devices\cite{Schiffer_surveymodelingmicrogrids_2016,Zhang_OnlineDynamicSecurity_2015,Simpson-Porco_VoltageStabilizationMicrogrids_2017}, and load with frequency and voltage response\cite{Kasis_PrimaryFrequencyRegulation_2017}. Examples of the generic model \eqref{eq:dynamicV} are presented in Section \ref{sec:control}.
\subsection{Modeling the Network}
Transmission lines are represented by the standard admittance matrix $Y=G+jB$. The power flow balance equations at each bus $i\in\mathcal{V}$ are given as follows.
\begin{equation}\label{eq:pf}
\begin{aligned}
P_i&=G_{ii}V_i^2+\sum_{j\in\mathcal{N}_i}V_iV_j(B_{ij}\sin\theta_{ij}+G_{ij}\cos\theta_{ij})\\
Q_i&=-B_{ii}V_i^2-\sum_{j\in\mathcal{N}_i}V_iV_j(B_{ij}\cos\theta_{ij}-G_{ij}\sin\theta_{ij})
\end{aligned}
\end{equation}
where $\mathcal{N}_i$ is the set of nodes who are adjacent to node $i$, $B_{ij}$ and $G_{ij}$ are elements in the admittance matrix.
For simplicity, we make the following assumption in this paper.
\begin{assumption}\label{as:1}
	The power network is lossless, i.e. the conductance matrix $G=\textbf{0}_{n\times n}$.
\end{assumption}
\begin{remark}
	\textit{
	Assumption \ref{as:1} is reasonable for the resistance is much smaller than the inductance in high-voltage transmission lines. And it is a common assumption in the context of power system stability research \cite{Chiang_Directstabilityanalysis_1995,Fiaz_portHamiltonianapproachpower_2013a,Stegink_unifyingenergybasedapproach_2016,Dorfler_Synchronizationcomplexoscillator_2013}.}
\end{remark}

Let $y=:(\theta_1,\ldots,\theta_n,V_1,\ldots,V_n)^T$ and $u=:(P_1,\ldots,P_n,$ $ Q_1,\ldots,Q_n)^T$. The network model \eqref{eq:pf} is a mapping from $y$ to $u$ and can be re-written in a compact form as follows.
\begin{equation}\label{eq:pfcompact}
u=g(y)
\end{equation}
\subsection{Modeling the Entire System}
Combining all bus dynamics \eqref{eq:dynamicV} and the network equation \eqref{eq:pfcompact}, the overall system can be formulated in a compact form as follows.
\begin{equation}\label{eq:entire}
\left\lbrace \begin{aligned}
\dot{x}&=f(x,u)\\
y&=h(x)\\
u&=g(y)
\end{aligned}\right. 
\end{equation}

\begin{figure}[h]
	\centering
	\setlength{\belowcaptionskip}{-0.5cm}
	\includegraphics[width=0.6\hsize]{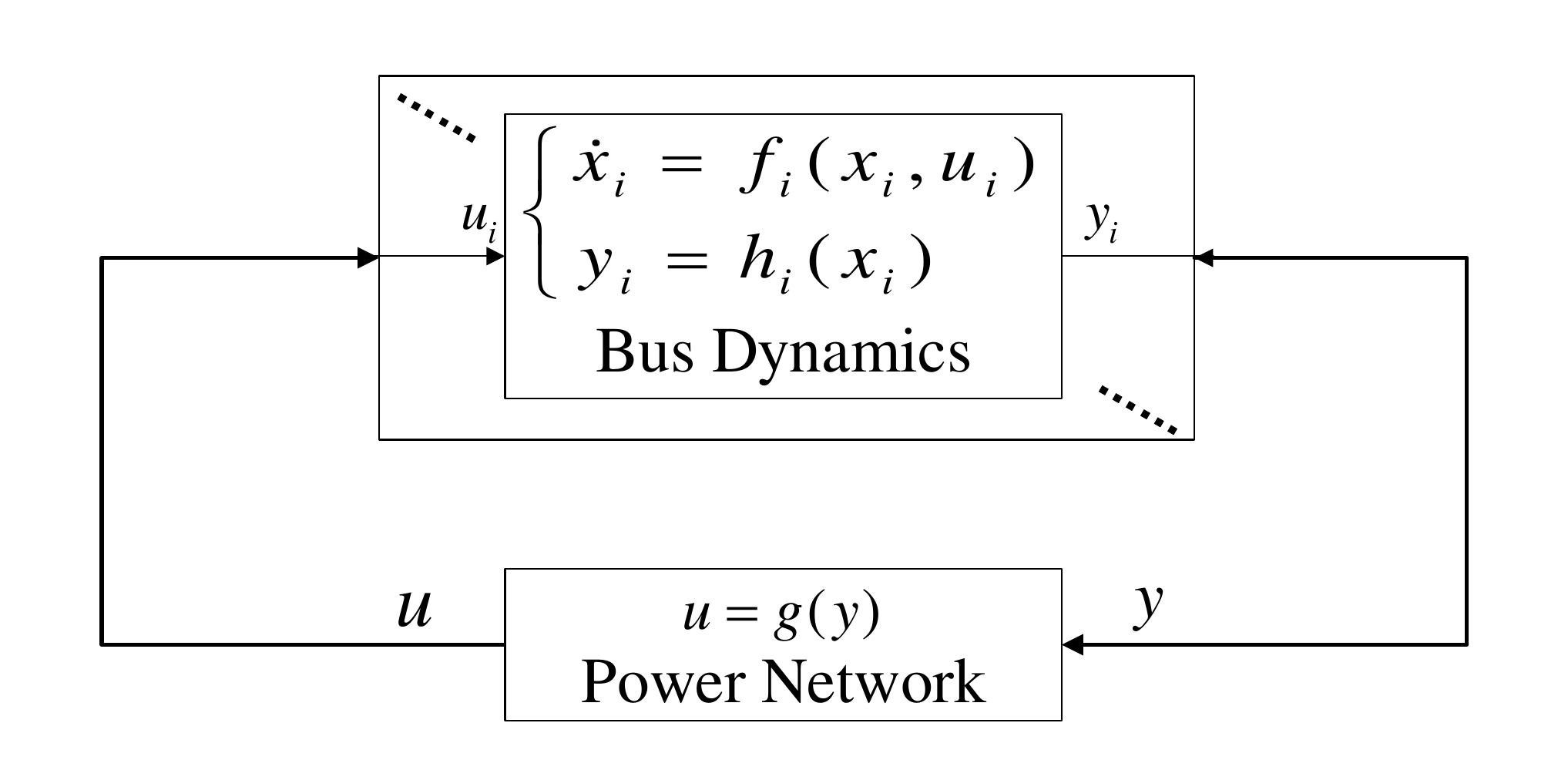}
	\caption{Input-output relation of the bus dynamics and the power network.}
	\label{fig:BusD}
\end{figure}

The overall system can be further represented as an input-output feedback interconnection between the network and every bus dynamics as shown in Fig. \ref{fig:BusD}.

\begin{definition}
	$x^*$ is called an equilibrium of system \eqref{eq:entire} if 
	\begin{equation*}
	\left\lbrace \begin{aligned}
	0&=f(x^*,u^*)\\
	y^*&=h(x^*)\\
	u^*&=g(y^*)
	\end{aligned}\right. 
	\end{equation*}
	Note that by \eqref{eq:dynamicV} and \eqref{eq:pf} the map from $x^*$ to $u^*,y^*$ is one-to-one. We denote by $(u^*,x^*,y^*)$  the input-state-output triplet associated with the equilibrium $x^*$.
\end{definition}

We now introduce the classical definitions of dissipativity and passivity index, which will be extended in the ensuing section. Consider a general nonlinear dynamical system $\Sigma: u\mapsto y$
\begin{equation}
\left\lbrace \begin{aligned}
\dot{x}&=f(x,u)\\
y&=h(x,u)
\end{aligned}\right. 
\end{equation}
where $x\in X$ and $u\in U$.

\begin{definition}[\textbf{Dissipativity and passivity \cite{Sepulchre_ConstructiveNonlinearControl_1997}}]\label{de:passive}
	System $\Sigma$ is said to be dissipative w.r.t. supply rate
	$w(t)$  if there exists a non-negative real continuously differentiable function $S(x)$, called the storage function, such that, for all $x_0\in X$
	and $u\in U$
	$$\dot{S}(x(t))\leq w(t)$$
	If $w(t)=u^Ty$, and $S(0)=0$, then $\Sigma$ is said to be passive.
\end{definition}
\begin{definition}[\textbf{Output feedback passivity \cite{Sepulchre_ConstructiveNonlinearControl_1997}}]\label{de:index}
	System 
	$\Sigma: u\mapsto y$ is said to be output feedback passive (OFP) if it is dissipative with respect to supply rate $w (u, y) = u^Ty-\sigma y^Ty$ for some $\sigma\in\rr$, denoted as OFP($\sigma$).
\end{definition}
The real number $\sigma$ quantifies the excess or shortage of passivity of system $\Sigma$ and is referred to as \emph{passivity index}. A positive $\sigma$ indicates excess of passivity while a negative $\sigma$ indicates shortage \cite{Sepulchre_ConstructiveNonlinearControl_1997}. In this paper, we extend this concept to analyze the passivity structure of power systems with interconnected heterogeneous bus dynamics.

\section{Distributed Stability Condition Induced by Passivity Index}
\label{sec:2}
In this section, we first specialized the supply rate $w$ with differential at one port. Then by quantifying the shortage of passivity of the power network, a distributed passivity index condition is derived for each bus dynamics to ensure the system-wide stability.

As the equilibrium of power system \eqref{eq:entire} is generally non-zero, we define the incremental supply rate with respect to an (arbitrary) input-state-output triplet $(u^*,x^*,y^*)$ of \eqref{eq:entire} as follows.
\begin{equation}\label{eq:sr}
w(u_i,y_i,\dot{y}_i)=-(P_i-P_i^*)\dot{\theta}_i-(\frac{Q_i}{V_i}-\frac{Q_i^*}{V_i^*})\dot{V}_i
\end{equation}
In terms of input-output pair, $P_i-P_i^*$ and $\frac{Q_i}{V_i}-\frac{Q_i^*}{V_i^*}$ are incremental values of input signals, and $(\dot{\theta}_i,\dot{V}_i)$ are derivatives as the output signals.
\begin{remark}
	\textit{
	Compared with the classical passivity framework, the supply rate \eqref{eq:sr} is a function of not only $u,y$ but also $\dot{y}$. This form of supply rate is related to the Brayton-Moser formulation \cite{Jeltsema_Multidomainmodelingnonlinear_2009a,Yang_DistributedStabilityAnalytics_2019} and the power shaping control technique \cite{Ortega_Powershapingnew_2003}. Differentials at one port rise up when the power, instead of the energy, perspective is adopted for modeling and analysis. The supply rate \eqref{eq:sr} also appears in the energy flow analysis of power systems\cite{chen2013energy}.}
\end{remark}

\begin{figure}[h]
	\centering
	\setlength{\belowcaptionskip}{-0.5cm}
	\includegraphics[width=1\hsize]{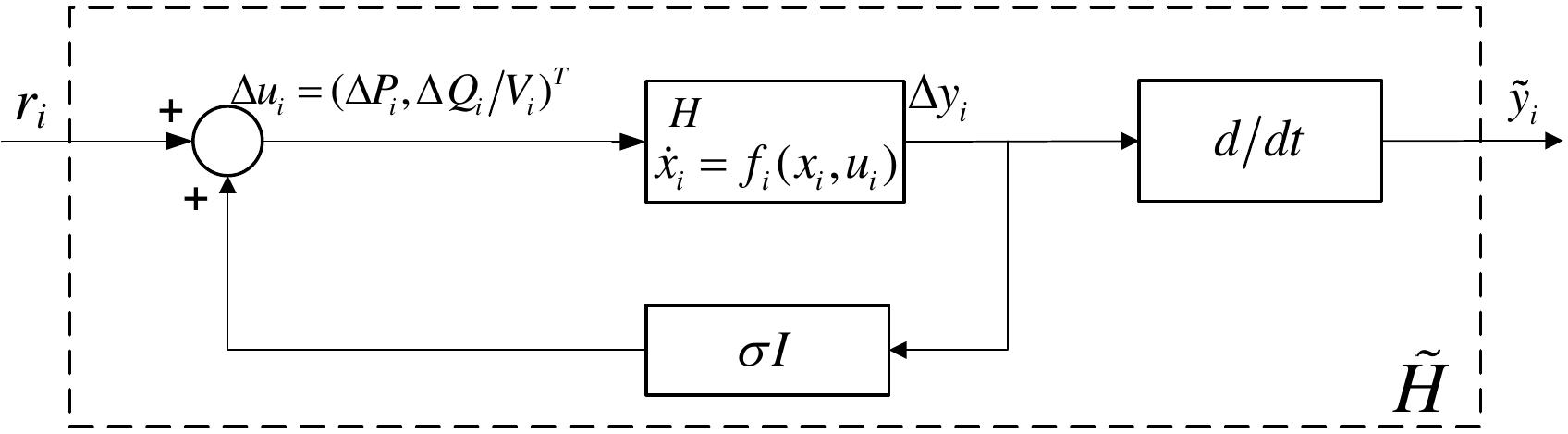}
	\caption{OFP($\sigma$) of $H$ is equivalent to standard passive of $\tilde{H}$.}
	\label{fig:Bus}
\end{figure}

\begin{definition}\label{de:OFD}
	Dynamical system \eqref{eq:dynamicV} is said to be OFP($\sigma$) w.r.t. an equilibrium  $x_i^*$ for some $\sigma\in\rr$, if there exist a real continuously differentiable storage function $S_i(x_i)$ such that
	\begin{equation}
	\dot{S}_i\leq-(P_i-P_i^*)\dot{\theta}_i-(\frac{Q_i}{V_i}-\frac{Q_i^*}{V_i^*})\dot{V}_i-\sigma(y_i-y_i^*)^T\dot{y}_i
	\end{equation}
\end{definition}
In terms of Definition \ref{de:OFD}, a dynamical system $H$ is OFP($\sigma$) if and only if the output feedback system $\tilde{H}$ with differential at one port, as shown in Fig.\ref{fig:Bus}, is passive in terms of Definition \ref{de:passive} taking $r_i$ and $\tilde{y}_i$ as input and output.
\subsection{Passivity Index of the Power Network}
Consider the following function $W_N:\rr^{n}\times\rr_{>0}^n\to\rr$
\begin{equation}
W_N(y)=\sum_{i\in\mathcal{V}}-\frac{1}{2}B_{ii}V_i^2-\sum_{(i,j)\in\mathcal{E}}B_{ij}V_iV_j\cos\theta_{ij}
\end{equation}

All eigenvalues of its Hessian matrix $\nabla^2W_N(y)$ are real since it is real and symmetric. $\nabla^2W_N(y)$ has one zero eigenvalue with eigenvector $\text{col}(\mathbf{1}_n,\mathbf{0}_n)$, which is caused by the \textit{rotational symmetry} of phase angle $\theta$ \cite{Dorfler_Synchronizationcomplexoscillator_2013}. Denote the minimal non-zero eigenvalue of $\nabla^2W_N(y^*)$ by $\lambda$. We will show that $\lambda$ is the passivity index of the power network.

For any input-state-output triplet $(u^*,x^*,y^*)$ of \eqref{eq:entire}, we choose the storage function for the network as follows.
\begin{equation}\label{eq:SN}
S_N(y)=\tilde{W}_N(y)-(\frac{\lambda-\varepsilon}{2})(y-y^*)^T(y-y^*)
\end{equation}
where $\varepsilon>0$ is any positive number and 
$$\tilde{W}_N(y)=W_N(y)-(y-y^*)^T\nabla W_N(y^*)-W_N(y^*)$$
\begin{lemma}\label{le:1}
	In some neighborhood of $y^*$, the network storage function $S_N$ satisfies $S_N(y)>0,\,\forall y\neq y^*$, $S_N(y^*)=0$ and 
	\begin{equation}\label{eq:le1}
	\dot{S}_N=\begin{bmatrix}
	P-P^*\\ \frac{Q}{V}-\frac{Q^*}{V^*}
	\end{bmatrix}^T\dot{y}-(\lambda-\varepsilon)(y-y^*)^T\dot{y}
	\end{equation}
\end{lemma}
The local minimum claim can be proved by the first and second derivative test and \eqref{eq:le1} is obtained by direct calculation, which is omitted here due to space limit.
\begin{remark}
	\textit{
	Letting $\varepsilon\to0$, Lemma \ref{le:1} indicates that $\lambda$ is the passivity index of the power network with a positive definite storage function. Depending on the operation point, $\lambda$ may be negative or positive, which indicates the network has either shortage or excess of passivity. Our simulation results suggest that the power network suffers from shortage of passivity in most cases while it may have excess of passivity only when the power flow is extremely light.}
\end{remark}

\subsection{Passivity Index Condition for System-Wide Stability}
To guarantee the system-wide stability of equilibrium $x^*$, we propose a distributed condition for each bus dynamics, which relies on the input-output property rather than specific dynamical models, such that the heterogeneity of bus dynamics can be dealt with in unified framework. For each bus dynamics, consider the following condition:\\
\textbf{Condition C1.} The bus dynamics \eqref{eq:dynamicV} is OFP($\sigma$) in term of Definition \ref{de:OFD} for some $\sigma>-\lambda$. And the storage function $S_i(x_i)$ has a strict local minimum at the equilibrium  $x_i^*$.

\begin{theorem}\label{th:1}
	For any equilibrium $x^*$ of \eqref{eq:entire}, if each bus dynamics satisfies Condition C1 then the equilibrium $x^*$ is asymptotically stable. 
\end{theorem}
%
%
%
Theorem \ref{th:1} can be proved by justifying $W(x)=\sum_{i\in\mathcal{V}}S_i(x_i)+S_N(y)+\frac{\sigma+\lambda-\varepsilon}{2}\left\| y-y^*\right\|^2$ as a Lyapunov function, which is omitted here due to space limit.
\begin{remark}
	\textit{
	When the network has deficit passivity, i.e. $\lambda<0$, Theorem \ref{th:1} says that if each bus dynamics is capable of compensating the shortage caused by the network, i.e. $\sigma+\lambda>0$, the interconnected system is ensured to be stable. From the perspective of interconnection, the excess of passivity of each bus dynamics percolates into the network via the electrical interconnection, enforcing the entire system to be passive, hence the system-wide stability is ensured. Theorem \ref{th:1} bridges the gap between component-wise passivity and system-wise stability, enabling a sufficient condition to justify the system stability in a distributed manner. }
\end{remark}

\section{Passivity Percolation via Control}\label{sec:control}
%
%
%
In this section, we give three typical examples of the generic model \eqref{eq:dynamicV}, and propose proper control designs such that the bus dynamics is equipped with enough passivity and Condition C1 is met.
\subsection{Synchronous Generators}
A typical power device of model \eqref{eq:dynamicV} is the flux-decay model of SG as follows\cite{Stegink_unifyingenergybasedapproach_2016}.
\begin{equation}\label{eq:sg1}
\left\lbrace 
\begin{aligned}
\dot{\delta}_i&=\omega_i\\
M_i\dot{\omega}_i&=-D_i\omega_i-P_i+P^g_i\\
T_{di}'\dot{E}_{qi}'&=-E_{qi}'-\frac{x_{di}-x_{di}'}{E_{qi}'}Q_i+E_{fi}
\end{aligned}\right. 
\end{equation}
where $E_{qi}'\angle\delta_i$ is the q-axis transient internal voltage. $\omega_i$ is the frequency derivation. $M_i$ is the moment of inertia. $D_i$ is the damping coefficient. $T_{di}'$ is the q-axis open-circuit transient time constant. $x_{di}$ and $x_{di}'$ are the d-axis synchronous reactance and transient reactance, respectively. For a realistic SG, $x_{di}'>x_{di}$. $P_i^g$ and $E_{fi}$ are control signals, which stand for the power generation and the excitation voltage, respectively. $P_i$ and $Q_i$ are the output active and reactive power, respectively.

For any given $\sigma\in\rr$, we propose the following controller to render SG \eqref{eq:sg1} OFP($\sigma$) in terms of Definition \ref{de:OFD} so that Condition C1 is satisfied.
\begin{proposition}\label{pro:sg}
	Set the control of \eqref{eq:sg1} as
	\begin{subequations}\label{eq:sgcontrol}
		\begin{equation}\label{eq:sgPg}
			P_i^g=-K_I\int_{0}^{\tau}\omega_idt-K_P\omega_i+P_i^{g*}
		\end{equation}
		\begin{equation}\label{eq:sgEf}
			E_{fi}=-K_E(E_{qi}'-E_{qi}'^*)+E_{fi}^*
		\end{equation}
	\end{subequations}
	where $K_I>\sigma$, $K_P>0$ and $K_E>(x_{di}-x_{di}')\sigma-1$ are constant. $P_i^{g*}$ and $E_{fi}^*$ are the steady-state inputs. Then the bus dynamics \eqref{eq:sg1} satisfies Condition C1.
\end{proposition}
It can be proved by setting the storage function as $S_i(x_i)=\frac{1}{2}M_i\omega_i^2+\frac{K_I-\sigma}{2}(\delta_i-\delta_i^*)^2	+\frac{1}{2}(\frac{K_E+1}{x_{di}-x_{di}'}-\sigma)(E_{qi}'-E_{qi}'^*)^2$ and calculate its derivative, which is omitted here.
\begin{remark}
	\textit{
	Note that \eqref{eq:sgPg} is a standard PI controller with the frequency derivation $\omega_i$ as input, and \eqref{eq:sgEf} is simply a negative feedback. This simple and classical controller is enough to meet Condition C1, which shows a promising potential of our condition for practical applications.}
\end{remark}

\subsection{Conventional Droop-Controlled Inverters}
Another typical bus dynamics of model \eqref{eq:dynamicV} is the inverter-interfaced device with conventional $P-\theta$ and $Q-V$ droop control as follows\cite{Zhang_OnlineDynamicSecurity_2015}.
\begin{equation}\label{eq:egV}
\left\lbrace 
\begin{aligned}
\tau_{i1}\dot{\theta}_i&=-(\theta_i-\theta_i^*)-D_{i1}(P_i-P_i^*)\\
\tau_{i2}\dot{V}_i&=-(V_i-V_i^*)-D_{i2}(Q_i-Q_i^*)
\end{aligned}\right. 
\end{equation}
where $\tau_{i1}$, $\tau_{i2}$ are the time constants and $D_{i1}$, $D_{i2}$ are the droop gains. Note that \eqref{eq:egV} could be either a source or a load depending on the sign of $P_i^*$ and $Q_i^*$.

For this kind of dynamics, we propose a condition on droop gains such that bus dynamics \eqref{eq:egV} is OFP($\sigma$) and Condition C1 is met.
\begin{proposition}\label{pro:CD}
	Let the droop gains $D_{i1}$ and $D_{i2}$ in \eqref{eq:egV} satisfy
	\begin{equation}\label{eq:conventionD}
	\begin{aligned}
	&D_{i1}^{-1}>\sigma,\\
	&D_{i2}^{-1}>(V_i^{*2}\sigma-Q_i^*)/V_i^*
	\end{aligned}
	\end{equation}
Then the bus dynamics \eqref{eq:egV} satisfies Condition C1.
\end{proposition}
%
This proposition can be proved by setting  the storage function as $S_i(x_i)=\frac{D_{i1}^{-1}-\sigma}{2}(\theta_i-\theta_i^*)^2+\frac{k_i}{D_{i2}}(\frac{V_i}{V_i^*}-\ln V_i)-\frac{\sigma(V_i-V_i^*)^2}{2}$, which is omitted here due to space limit.
\subsection{Quadratic Droop-Controlled Inverters}
We now consider another typical bus dynamics in the literature, which is known as the quadratic droop controller\cite{Simpson-Porco_VoltageStabilizationMicrogrids_2017}. Its dynamics can be expressed as follows.
\begin{equation}\label{eq:egQD}
\left\lbrace 
\begin{aligned}
\tau_{i1}\dot{\theta}_i&=-(\theta_i-\theta_i^*)-D_{i1}(P_i-P_i^*)\\
\tau_{i2}\dot{V}_i&=-D_{i2}Q_i-V_i(V_i-u_i^*)
\end{aligned}\right. 
\end{equation}
where $u_i^*$ is a constant satisfying
\begin{equation}\label{eq:QDu}
0=-D_{i2}Q_i^*-V_i^*(V_i^*-u_i^*)
\end{equation}

For this kind of bus dynamics, we have the following proposition for droop gain settings.
\begin{proposition}\label{pro:QD}
	Let the droop gains $D_{i1}$ and $D_{i2}$ in \eqref{eq:egQD} satisfy
	\begin{equation}\label{eq:QDroop}
	D_{i1}^{-1}>\sigma,\quad D_{i2}^{-1}>\sigma
	\end{equation}
	Then the bus dynamics \eqref{eq:egQD} satisfies Condition C1.
\end{proposition}
%
It can be proved by setting the storage function as $S_i(x_i)=\frac{D_{i1}^{-1}-\sigma}{2}(\theta_i-\theta_i^*)^2+\frac{D_{i2}^{-1}-\sigma}{2}(V_i-V_i^*)^2$, which is omitted here. 
\begin{remark}
	\textit{
	Proposition \ref{pro:CD} and \ref{pro:QD} both require the droop gains less than certain values, which is consistent with the literature that a too large  gain can cause instability \cite{Pogaku_ModelingAnalysisTesting_2007b}.}
\end{remark}
\section{Case Study}
\begin{figure}[h]
	\centering
	\includegraphics[width=0.55\hsize]{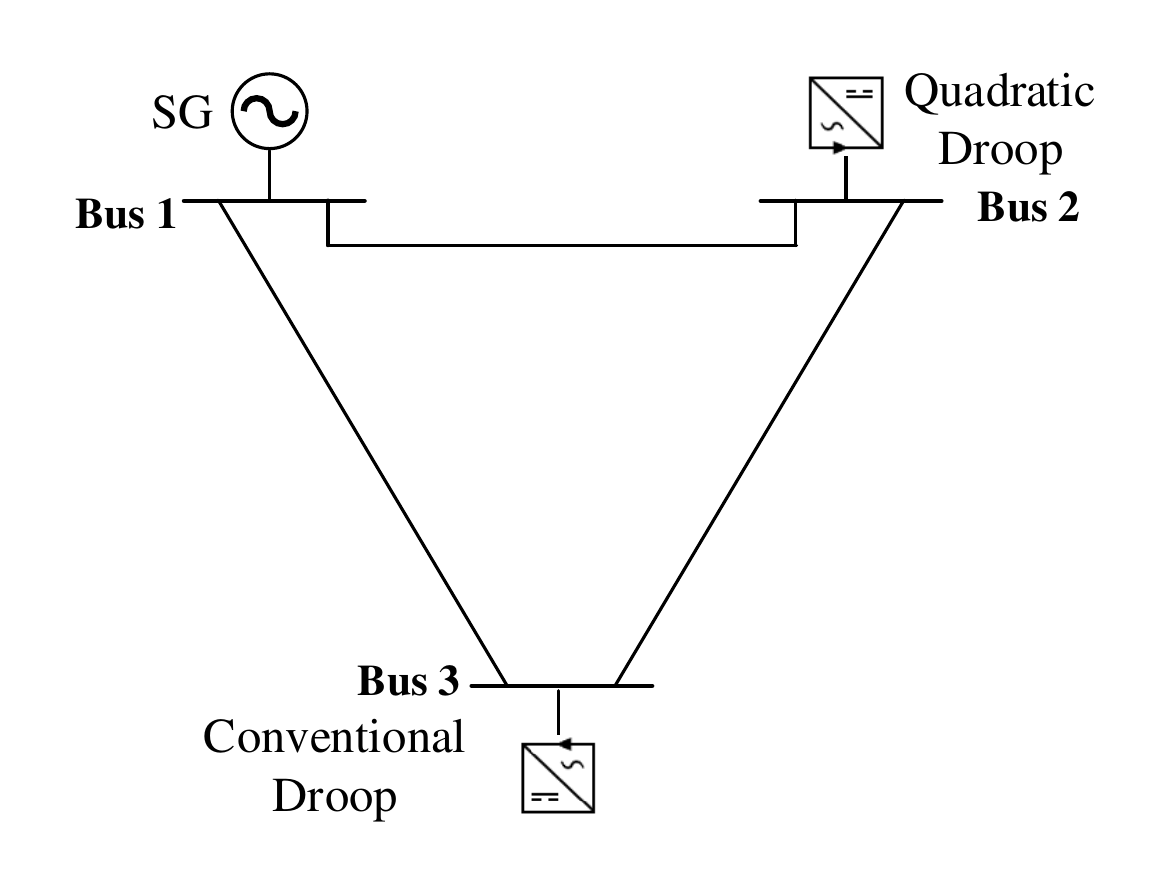}
	\caption{The schematic of the 3-bus power system with different bus dynamics.}
	\label{fig:3bus}
\end{figure}

Consider a 3-bus power system as shown in Figure~\ref{fig:3bus}. Bus 1 is connected with a SG \eqref{eq:sg1}\eqref{eq:sgcontrol}. Bus 2 is attached to a quadratic droop-controlled inverter \eqref{eq:egQD}. And Bus 3 is connected with a conventional droop-controlled inverter \eqref{eq:egV} as a load. The parameters of the system are listed in Table~\ref{tab1}.
\vspace{-0.2in}

\begin{table}[h]
	\centering
	\footnotesize
	\caption{System parameters}
	\label{tab1}
	\begin{tabular}{l|l}
		\hline
		Parameters&Values (p.u.)\\
		\hline
		Transmission Lines $x$ $r$        & 0.12, 0.01  \\ 
		SG parameters $M_i,D_i,T_{di}',x_{di},x_{di}'$   & 0.16, 0.076, 6.56, 0.295, 0.17 \\ 
		QD parameters $\tau_{i1},\tau_{i2}$ & 0.3, 8 \\ 
		CD parameters $\tau_{i1},\tau_{i2}$& 1, 10 \\
		\hline
	\end{tabular}
\end{table}
\subsection{Passivity Index of the Network}
\begin{figure}[h]
	\centering
	\includegraphics[width=0.55\hsize]{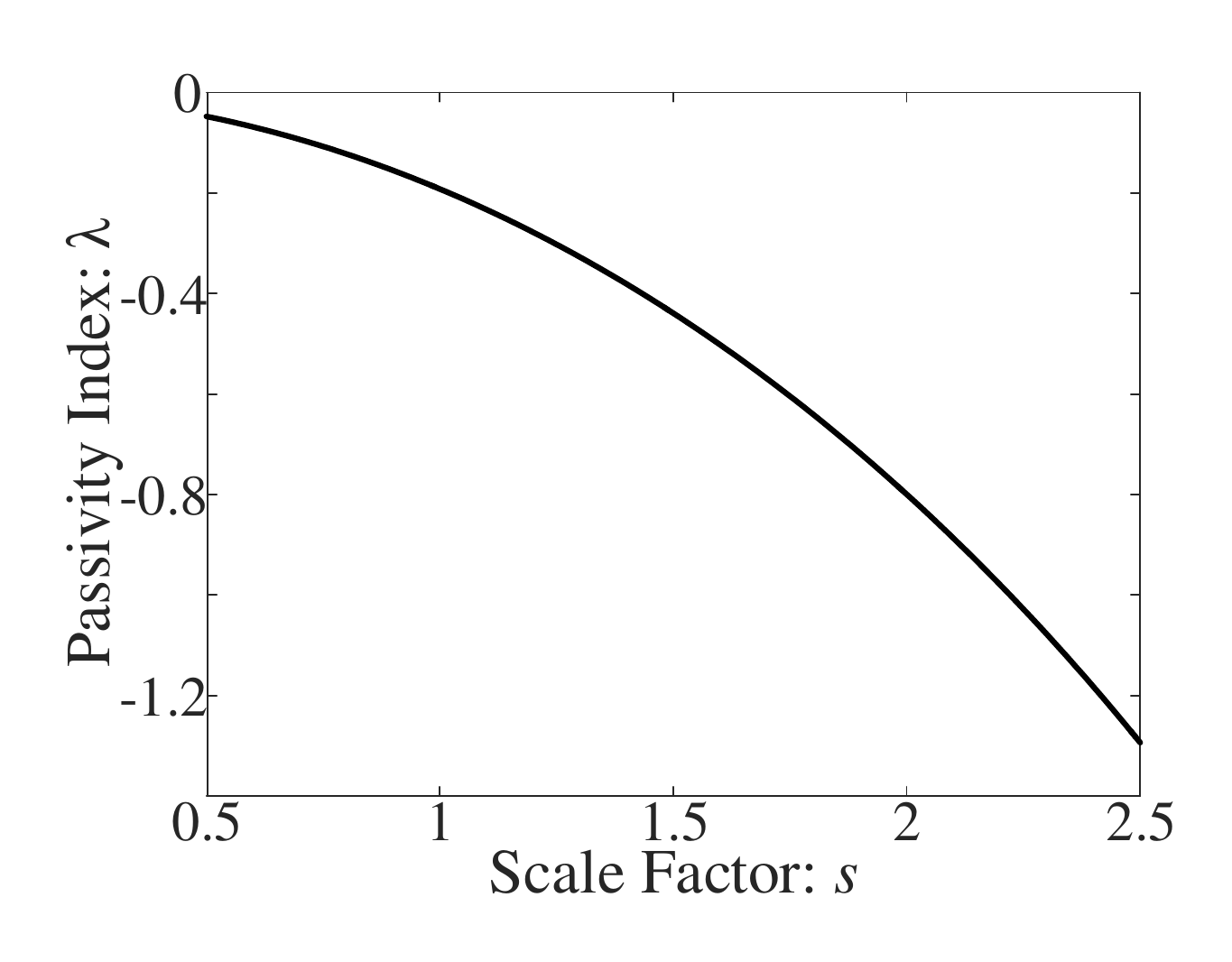}
	\caption{The passivity index $\lambda$ varies with load profile.}
	\label{fig:lamda}
\end{figure}

To solve the power flow and obtain the system equilibrium, we set bus 1 as the $V\theta$ node, bus 2 as a $PV$ node, and bus 3 as a $PQ$ node. The base load profile is $P_2=1,P_3=-1.5,Q_3=-0.1$. In order to the show the characteristics of $\lambda$ changing with the load, the base profile is multiplied by a scale factor $s$, from 0.5 to 2.5, to simulate different load conditions. The corresponding passivity index $\lambda$ is calculated for each $s$. The result is shown in Fig. \ref{fig:lamda}. 
\vspace{-0.05in}

Fig. \ref{fig:lamda} shows that the power network has shortage of passivity and its index decreases as the loads scale up. It implies that the stability margin deteriorates as loads increase, and bus dynamics should support more passivity to maintain system-wide stability.

\subsection{Small-Signal Stability Verification and Tightness}
\begin{figure}[h]
	\footnotesize
	\centering
	\setlength{\abovecaptionskip}{0.cm}	
	\setlength{\belowcaptionskip}{-0.cm}
	\includegraphics[width=1\columnwidth]{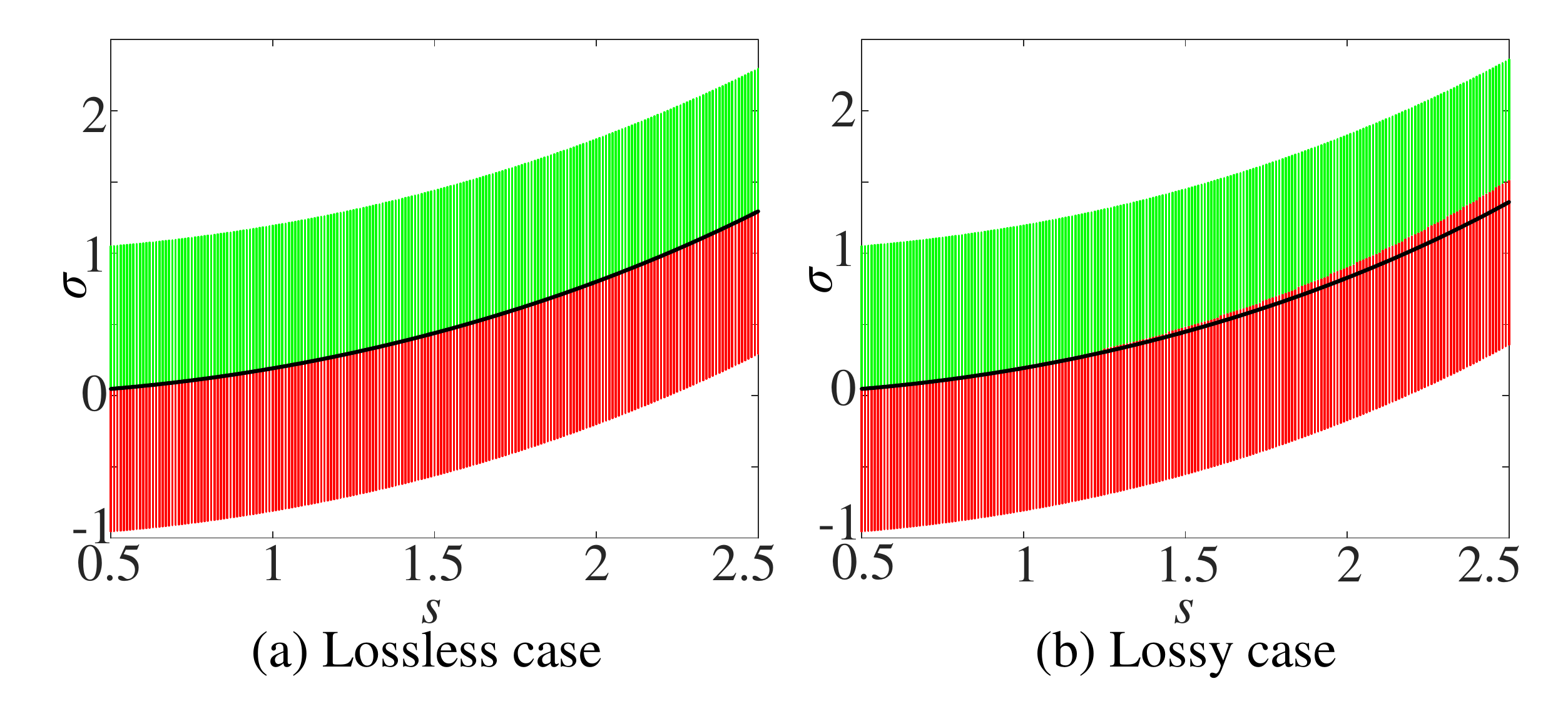}
	\caption{The stability verification for (a) the lossless case and (b) the lossy case. The minimal passivity index required by our condition, namely $-\lambda$, in different load scales is marked by the black line. For each $s$, $\sigma$ varies around $-\lambda$, resulting in system-wide small-signal stability (marked in green) or instability (marked in red).}
	\label{fig:tight_lossless&lossy}
\end{figure}

We set the controls according to Proposition \ref{pro:sg}-\ref{pro:QD} such that every bus dynamics is OFP($\sigma$). Theorem \ref{th:1} affirms that if every bus satisfies $\sigma>-\lambda$ then the system-wide stability is guaranteed. To verify this claim and illustrate the tightness, for each scale factor $s$, we set $\sigma=-\lambda+\rho$ where $\rho$ rangers from $-1$ to $1$. Then, for each pair of $(s,\sigma)$ we calculate eigenvalues of the system-wide Jacobian matrix. If all eigenvalues have negative real parts, the system is small-signal stable and the point $(s,\sigma)$ is colored in green. Otherwise, the system is unstable and the point $(s,\sigma)$ is colored in red. The results are shown in Fig. \ref{fig:tight_lossless&lossy} for both lossless and lossy cases.

It is clear, in the lossless situation, the system-wide stability is always guaranteed if every bus obeys $\sigma>-\lambda$, which justifies Theorem \ref{th:1}. Note that the upper fringe of the red area indicates the exact minimal $\sigma$ to ensure system-wide stability. Fig. \ref{fig:tight_lossless&lossy}(a) suggests our condition is almost necessary for lossless case. It is worth mentioning that when the $\sigma$ of one/two bus dynamics is fixed at the marginal value $-\lambda$, and the other two/one buses vary $\sigma=-\lambda+\rho$, the same result appears as Fig. \ref{fig:tight_lossless&lossy}(a), which means any violation of Condition C1 could cause instability.

A slightly different result arises in the lossy case as shown in Fig. \ref{fig:tight_lossless&lossy}(b), where a little larger $\sigma$ is in need to induce system-wide stability when the system is heavily loaded. It also implies the existence of transfer conductance may undermine the system stability.
\subsection{Transient Stability Tests}
In this subsection, we illustrate the relation between passivity index and the system-wide transient stability. Consider the base load profile, i.e. $s=1$. A three-phase short circuit fault occurs at Bus 1, 2, and 3, respectively. For each fault, we set the passivity index of bus dynamics at $\sigma=-\lambda$, $\sigma=-\lambda+1$, and $\sigma=-\lambda+2$ in turn. For each case, the critical clearing time (CCT) is calculated via numerical simulation. The results of lossless and lossy system are shown in Table \ref{tab:CCT} and \ref{tab:CCT2}, receptively.
\begin{table}[htb]
	\footnotesize
	\centering
	\caption{Lossless: CCT(s) in different cases}
	\label{tab:CCT}
	\begin{tabular}{c|c|c|c}
		\hline
		\multirow{2}{*}{Fault Bus} & \multicolumn{3}{|c}{Passivity Index $\sigma$}\\
		\cline{2-4}
		&$-\lambda$&$-\lambda+1$&$-\lambda+2$\\
		\hline
		 Bus 1&0.076 &0.205 &0.414 \\
		 Bus 2&0.090 &0.291 &0.625 \\
		 Bus 3&0.087 &0.287 & 0.618\\
		\hline
	\end{tabular}
\end{table}

\begin{table}[htb]
	\footnotesize
	\centering
	\caption{Lossy: CCT(s) in different cases}
	\label{tab:CCT2}
	\begin{tabular}{c|c|c|c}
		\hline
		\multirow{2}{*}{Fault Bus} & \multicolumn{3}{|c}{Passivity Index $\sigma$}\\
		\cline{2-4}
		&$-\lambda$&$-\lambda+1$&$-\lambda+2$\\
		\hline
		Bus 1&0.055 &0.175 &0.365 \\
		Bus 2&0.062 &0.243 &0.549 \\
		Bus 3&0.060 &0.241 & 0.545\\
		\hline
	\end{tabular}
\end{table}

It is clear that a larger $\sigma$ results in a larger CCT, which implies that the more passivity provided by bus dynamics, the more stable the system is.
\section{Concluding Remarks}
In this paper, we have derived a passivity index condition to guarantee the stability of lossless power systems with interconnected nonlinear heterogeneous bus dynamics. It is proved that if each bus dynamics is output feedback passive with a large enough passivity index, the system-wide stability can be ensured. Moreover, for three typical dynamical models, we have shown that the excess of passivity can be obtained via proper control settings. The theoretical result is verified by simulations of a rudimentary 3-bus power system. It is noticed that the proposed condition is quite tight and can still be valid when the resistance in transmission lines is considered. It also implies that a larger index is beneficial to transient stability.

The implication of this work includes a new approach to assess the system-wide stability in a distributed manner. The proposed condition is adaptable to nonlinear heterogeneous models, since it relies only on the input-output property. Moreover, as indicated in Section IV, the condition could serve as a scalable criterion and motivates controller designs for a diversity of existing models.

\bibliographystyle{IEEEtran}
\bibliography{mybib}

\end{document}